\documentclass[12pt]{article}
\usepackage{amsmath}
\usepackage{graphicx,subfigure}
\usepackage{latexsym}
\usepackage{amsfonts}
\usepackage{amsthm}
\newtheorem{thm}{Theorem}[section]

\title{New Complexiton Solutions of the KdV and Coupled KdV Equations}

\author {
Asl{\i} Pekcan \thanks{Email:aslipekcan@hacettepe.edu.tr}\\
{\small Department of Mathematics, Faculty of Science} \\
{\small Hacettepe University, 06800 Ankara - Turkey}}

\date{\nonumber}
\begin{document}
\maketitle
\date{\nonumber}

\numberwithin{equation}{section}

\begin{abstract}
 A new approach to double-sub equation method is introduced to construct novel solutions
for the nonlinear partial differential equations. It is
applied to the Korteweg-de Vries (KdV) equation and yields new complexiton solutions of
both the KdV and coupled KdV equations.
The graphs of the solutions are also illustrated.


\end{abstract}

\section{Introduction}

The Korteweg de Vries (KdV) equation
\begin{equation}\label{KdV}
u_t+\zeta uu_x+\rho u_{xxx}=0,\quad \quad \zeta, \rho\, \mathrm{arbitrary}\, \mathrm{constants}.
\end{equation}
 is a famous $(1+1)$-dimensional integrable nonlinear partial differential equation
which models waves on shallow water surfaces, ion acoustic waves in plasma, acoustic waves on a crystal lattice etc. It is
 introduced by Boussinesq \cite{Bo} and rediscovered by Korteweg and de-Vries \cite{KdV}. Since then there has been a huge interest
 on analysis of its rich structure- Lax pairs \cite{Lax}, symmetries \cite{Olver}, B\"{a}cklund transformations \cite{WahEst}, and so on, and finding its solutions. Researches on exploring the solutions inspired important methods like inverse scattering method \cite{GGKM}-\cite{AbCl}, Hirota direct method \cite{HirotaKdV}-\cite{hietarinta}, and Painlev\'{e} expansion \cite{WeissTaborGarnevale}-\cite{Sah}.

 The KdV equation has various type of solutions like soliton \cite{HirotaKdV}, and negaton, positon \cite{RSK}, and complexiton solutions \cite{MaComplexiton1}-\cite{MaComplexiton3}. There is a great interest to obtain new type of solutions hence several solution methods whose efficiency should also be discussed are constructed. One of them is the sub-ordinary differential equation method \cite{subODE1}-\cite{subODE3}. It was improved and had a new name 'double-sub equation method'  in \cite{chen-yang-ma}. The authors of \cite{chen-yang-ma} claimed that by using this method one can get complexiton solutions combining Jacobi elliptic functions and elementary functions. They applied this method to the KdV and mKdV equations. In \cite{liu-lin-sun}, it was shown that the solutions found in \cite{chen-yang-ma} are just constant solutions. But the method can still be used to find solutions of nonlinear partial differential equations.

\noindent In this paper, to get a wider class of solutions including complexiton solutions depending on two independent variables we use an ansatz different than the one in \cite{chen-yang-ma} for the form of the solution. In fact, we really obtain new complexiton solutions of the KdV equation.

\noindent Another set of the solutions yielded in this paper is for the coupled KdV equation. One type of the coupled KdV equations
\begin{equation}\label{coupledKdV}
\begin{aligned}
U_t+\zeta UU_x-\zeta VV_x+\rho U_{xxx}&=0\\
V_t+\zeta UV_x+\zeta VU_x+\rho V_{xxx}&=0
\end{aligned}
\end{equation}
can be obtained from the KdV equation by simply using the transformation $u(x,t)=U(x,t)+\.{i}V(x,t)$ and separating the real and imaginary parts.
There are many coupled KdV systems like the Hirota and Satsuma system \cite{HirotaSat}, the Drinfeld and Sokolov model \cite{DrinfeldSokolov},
the Fuchssteiner equation \cite{Fuch}, the Nutku-O\~{g}uz model \cite{NO}, the degenerate coupled KdV systems \cite{MGAP1}, \cite{MGAP2}, and so on. The coupled KdV system (\ref{coupledKdV}) used in this paper was derived as a special case of the general coupled KdV systems obtained from the two layer model of the atmospheric dynamical systems \cite{Hu1} and two-component Bose-Einstein condensates \cite{Bose}. Even the complexiton solutions have been studied for several years, in our knowledge, there was no non-singular complexiton solution until the work of Hu et al. \cite{Hu2}. They discovered non-singular complexiton solutions of (\ref{coupledKdV}) by using the iterative Darboux transformations. Afterwards, Yang and Mao in \cite{YangMao1} obtained non-singular complexiton solutions of (\ref{coupledKdV}) different from the solutions in \cite{Hu2} by applying Hirota's direct method and conjugate number form of exponential functions. In this paper, investigating solutions of the KdV equation by the new approach reveals novel non-singular complexiton solutions of the coupled KdV equation different than the other solutions obtained before \cite{Hu2}, \cite{YangMao1}.

This paper is organized as follows. In Section II, we introduce the new approach to double-sub equation method and apply it to the KdV equation. Solutions depending on one and two independent variables are given. The latter case is analyzed in detail. The complexiton solutions of the coupled KdV equation obtained during the application of the new approach to the KdV equation are also given. Graphs of the solutions are illustrated. In Section III, asymptotic behaviors of the solutions are discussed.

\section{A New Approach}

In \cite{chen-yang-ma} the authors presented a method called double-sub equation method, which we will call it as Ma's approach. They assume the form of
the solution as
 \begin{equation}\label{Masolnform}
u(x,t)=a_0+\frac{a_1F(\xi)+a_2G(\eta)}{\mu_0+\mu_1F(\xi)+\mu_2G(\eta)}+\frac{a_3F(\xi)^2+a_4F(\xi)G(\eta)+a_5G(\eta)^2}{(\mu_0+\mu_1F(\xi)+\mu_2G(\eta))^2},
\end{equation}
 where the functions $F(\xi)$ and $G(\eta)$ satisfies first order ODEs
 \begin{equation}\label{MaODEs}
(F')^2(\xi)=\alpha_1+\beta_1 F^2(\xi)+\gamma_1 F^4(\xi), \quad
(G')^2(\eta)=\alpha_2+\beta_2 G^2(\eta)
\end{equation}
 where $\xi=k_1(x-c_1t)$ and $\eta=k_2(x-c_2t)$ with unknown constants $a_i$, $i=0,1,\ldots,5$; $\mu_j$, $j=0,1,2$; $k_m$, $c_m$, $m=1, 2$ that will be determined. As noted before, the solutions in \cite{chen-yang-ma} are constants but indeed, this approach gives non-constant but not original solutions. Some of the non-constant solutions of the KdV equation which can be obtained by Ma's approach are the followings.

\noindent If $\displaystyle a_1=a_2=a_4=a_5=\mu_1=\mu_2=0, a_3=-12\rho k_1^2\mu_0^2\gamma_1/\zeta$, and $c_1=4\rho k_1^2\beta_1+a_0\zeta$, we obtain the solution
\begin{equation}
\displaystyle u(x,t)=a_0-\frac{24\rho \alpha_1\gamma_1 k_1^2}{\zeta\omega}\mathrm{sn}^2\Big(\frac{\sqrt{2\omega}}{2}\xi+\theta_1,k \Big),
\end{equation}
where $\xi=k_1(x-c_1t)$, $k=\sqrt{-2\gamma_1\alpha_1}/\sqrt{2\gamma_1\alpha_1+\beta_1\omega}$ with $\omega=-\beta_1 +\sqrt{\beta_1^2-4\gamma_1\alpha_1}$, $c_1=4\rho k_1^2\beta_1+a_0\zeta$, and $\theta_1$ is any constant. This solution can easily be obtained by using symmetry reduction.

\noindent Another solution occurs when $\displaystyle a_1=a_3=a_4=\mu_1=\alpha_2=0, a_0=(12c_2\mu_2-\zeta a_2)/12\zeta \mu_2, a_5=-a_2\mu_2$, and $\displaystyle \beta_2=\zeta a_2/12\rho k_2^2\mu_2$. We get
\begin{equation*}\displaystyle
u(x,t)=\frac{12c_2\mu_2-\zeta a_2}{12\zeta\mu_2}+\frac{a_2 e^{\pm \frac{\sqrt{\zeta a_2}}{2k_2\sqrt{3\rho\mu_2}}(\eta+\theta_1) }}{\mu_0+\mu_2 e^{\pm \frac{\sqrt{\zeta a_2}}{2k_2\sqrt{3\rho\mu_2}}(\eta+\theta_1)}}-\frac{a_2\mu_2 e^{\pm \frac{\sqrt{\zeta a_2}}{k_2\sqrt{3\rho\mu_2}}(\eta+\theta_1) }}{\Big(\mu_0+\mu_2 e^{\pm \frac{\sqrt{\zeta a_2}}{2k_2\sqrt{3\rho\mu_2}}(\eta+\theta_1)}\Big)^2},
\end{equation*}
where $\eta=k_2(x-c_2 t)$ and $\theta_1$ is any constant.

\noindent Now we explain our approach to double-sub equation method which produces novel complexiton solutions depending on two independent variables and also periodic and solitary wave solutions. Let
\begin{equation}\label{anypde}
\Lambda (u, u_x, u_t, u_{xx}, ...)=0
\end{equation}
be a partial differential equation in two independent variables. Assume the following ansatz for the solutions of (\ref{anypde})
\begin{equation}\label{transformation}
u(x,t)=a_0+\frac{\kappa_1+a_1F(\xi)+a_2G(\eta)}{\mu_0+\mu_1F'(\xi)+\mu_2G'(\eta)}+\frac{\kappa_2+a_3F(\xi)^2+a_4F(\xi)G(\eta)+a_5G(\eta)^2}{(\mu_0+\mu_1F'(\xi)+\mu_2G'(\eta))^2},
\end{equation}
where $\mu_i$, $i=0, 1, 2$; $a_j$, $j=0, \ldots, 5$, $\kappa_m$, $m=1, 2$ are unknown constants. Here the functions $F(\xi)$ and $G(\eta)$ satisfy the following ODEs,
\begin{equation}\label{ODEs}
(F')^2(\xi)=\alpha_1+\beta_1 F^2(\xi)+\gamma_1 F^4(\xi),\quad
(G')^2(\eta)=\alpha_2+\beta_2 G^2(\eta)+\gamma_2 G^4(\eta)
\end{equation}
with $\xi=k_1(x-c_1 t)$ and $\eta=k_2(x-c_2 t)$, and $\alpha_i$, $\beta_i$, $\gamma_i$, $k_i$, $c_i$, $i=1,2$ are unknown constants to be determined.
The ansatz for the form of the solution used here is different than (\ref{Masolnform}) so that it gives wider class of solutions including the solutions depending on two independent variables $\xi$ and $\eta$. Substituting
(\ref{transformation}) into (\ref{KdV}) and using the constraints (\ref{ODEs}) give a system of
equations with respect to $F^{i}G^{j}F'^{r}G'^{p}$, $0\leq i, j \leq 10$; $r, p= 0,1$. We set the coefficients of $F^{i}G^{j}F'^{r}G'^{p}$, $0\leq i, j \leq 10$; $r, p= 0,1$ to vanish. This yields a set of equations in terms of the unknowns, $a_i$, $i=0,1 , \ldots, 5$; $\mu_j$, $j=0,1,2$; $\kappa_l$, $c_l$, $k_l$, $\alpha_l$, $\beta_l$, and $\gamma_l$, $l=1,2$. By the help of MAPLE, we solve these equations and obtain solutions depending on one independent variable and two independent variables. Indeed, the latter case is more interesting and there are not so many such kind of solutions. Therefore in the next section we will just present two examples related to the first case then focus on the solutions having two independent variables.

\subsection{Solutions depending on one independent variable}

Some solutions of the KdV equation depending on one independent variable obtained by the new approach are the followings.

\noindent Let $a_1=a_2=a_4=a_5=\kappa_1=\kappa_2=\mu_0=\mu_2=0$, $\displaystyle a_3=(48\rho\alpha_1\gamma_1\mu_1^2k_1^2-12\rho\beta_1^2\mu_1^2k_1^2)/\zeta$, and $c_1=-8\rho \beta_1 k_1^2+a_0\zeta$. Then the solution is
\begin{equation*}
\displaystyle u(x,t)=a_0+\frac{48\rho\mu_1^2k_1^2(4\gamma_1\alpha_1-\beta_1^2)\mathrm{sn}^2(\delta\xi+A_1, k)}{\mu_1^2
(-2\beta_1+2\sqrt{\beta_1^2-4\gamma_1\alpha_1})\mathrm{cn}^2(\delta\xi+A_1, k )\mathrm{dn}^2(\delta\xi+A_1, k) },
\end{equation*}
where
\begin{equation*}
\displaystyle \delta=\frac{1}{2}\sqrt{-2\beta_1+2\sqrt{\beta_1^2-4\gamma_1\alpha_1}},\quad \displaystyle k=\frac{\sqrt{-2(2\gamma_1\alpha_1-\beta_1^2+\beta_1\sqrt{\beta_1^2-4\gamma_1\alpha_1 })\gamma_1\alpha_1}}{2\gamma_1\alpha_1-\beta_1^2+\beta_1\sqrt{\beta_1^2-4\gamma_1\alpha_1}},
\end{equation*}
and $A_1$ is an arbitrary constant. This solution is a periodic solution.

\bigskip

\noindent Now let us take the parameters as $a_1=a_2=\kappa_1=\kappa_2=\mu_0=\gamma_1=\gamma_2=0$,
\begin{eqnarray}\label{case1cond}
&&a_3=-\frac{4\rho\mu_1^2\beta_1^2k_1^2}{3\zeta},\quad  a_4=\frac{8\rho\mu_1\beta_1^2k_1^3\mu_2}{9k_2\zeta},\quad a_5=-\frac{4\rho\mu_2^2\beta_1^2k_1^4}{27\zeta k_2^2}
\nonumber \\&&\alpha_2=\frac{\alpha_1 \mu_1^2}{\mu_2^2}, \quad \beta_2=\frac{\beta_1k_1^2}{9k_2^2},\quad c_2=3c_1+\frac{16}{9}\rho\beta_1 k_1^2-2\zeta a_0.
\end{eqnarray}
Choosing $\alpha_1>0$ and $\beta_1>0$, we obtain the following non-constant functions $F$ and $G$ from (\ref{ODEs}),
\begin{equation*}
\displaystyle F(\xi)=\pm \sqrt{\frac{\alpha_1}{\beta_1}}\sinh (\sqrt{\beta_1}(\xi+A_1)),\quad G(\eta)= \pm \sqrt{\frac{\alpha_2}{\beta_2}}\sinh (\sqrt{\beta_2}(\eta+A_2)),
\end{equation*}
where $A_1$, $A_2$ are arbitrary constants. By assuming $\mu_1\mu_2>0$ and $k_1k_2>0$, we have the solution
\begin{equation}\label{soln2}
u(x,t)=a_0-\frac{4\rho k_1^2\beta_1}{3\zeta}\Big[\frac{\sinh(\tau_1)-\sinh(\tau_2)}{\cosh(\tau_1)+\cosh(\tau_2)}\Big]^2,
\end{equation}
with $\tau_1=\sqrt{\beta_1}k_1(x-c_1t+\tilde{A}_1)$, $\displaystyle \tau_2=\sqrt{\beta_1}k_1(x-c_2t+\tilde{A}_2)/3$ where $\tilde{A}_1$ and $\tilde{A}_2$ are arbitrary constants. Even the solution seems depending on two independent variables, by using simple identities we can combine the hyperbolic functions and get a new function depending on one variable that is
\begin{equation}
\displaystyle u(x,t)=a_0-\frac{4\rho\beta_1k_1^2}{3\zeta}\tanh^2(k_3x-c_3t+A_3),
\end{equation}
 where $\displaystyle k_3=\sqrt{\beta_1}k_1/3$, $\displaystyle c_3=\sqrt{\beta_1}k_1(9\zeta a_0-8\rho\beta_1k_1^2)/27$, and $A_3$ is an arbitrary constant. This solution is clearly a solitary wave solution.
\newpage
\subsection{Solutions depending on two independent variables}

\noindent Here we consider the solutions that include both of the functions $F(\xi)$ and $G(\eta)$.
\begin{thm}
The KdV equation has a solution of the form (\ref{transformation}) with the relations (\ref{ODEs}) satisfied, depending on two independent variables if and only if $\gamma_1=\gamma_2=0$.
\end{thm}

\noindent We present all the solutions of the KdV equation having two independent variables found by the new approach.

\noindent \textbf{Case 1.} Let $\gamma_1=\gamma_2=\mu_0=a_1=a_2=\kappa_1=0$,
\begin{eqnarray*}
&& a_3=-\frac{252k_2^4\beta_2^2\rho\mu_1^2}{k_1^2\zeta}, \quad a_4=\frac{168\rho\mu_1\mu_2k_2^3\beta_2^2}{k_1\zeta}, \quad  a_5=\frac{36\rho\mu_2^2\beta_2^2k_2^2}{\zeta}, \quad \kappa_2=\frac{384\rho\alpha_2k_2^2\mu_2^2\beta_2}{7\zeta},\\
&&\alpha_1=-\frac{\alpha_2 \mu_2^2}{7\mu_1^2}, \quad
\beta_1=-\frac{7\beta_2 k_2^2}{k_1^2},\quad c_1=32\rho \beta_2 k_2^2+\zeta a_0,\quad c_2=16\rho \beta_2 k_2^2+\zeta a_0.
\end{eqnarray*}

\noindent \textbf{i)} \textbf{$\alpha_2<0$, $\beta_2>0$}. The only case that we have real-valued solutions is when $\alpha_2<0$ and $\beta_2>0$. In this case, we obtain the functions $F(\xi)$ and $G(\eta)$ from (\ref{ODEs}) as
\begin{equation*}
\displaystyle F(\xi)=\pm \frac{\sqrt{-\alpha_2}k_1\mu_2}{7\mu_1k_2\sqrt{\beta_2}}\sin\Big(\frac{\sqrt{7\beta_2}k_2}{k_1}(\xi+A_1)\Big),  \quad
G(\eta)=\pm \frac{\sqrt{-\alpha_2}}{\sqrt{\beta_2}}\cosh(\sqrt{\beta_2}(\eta+A_2)),
\end{equation*}
so the solution is
\begin{equation}\label{case1soln}
u(x,t)=a_0+\frac{A}{B},
\end{equation}
where
\begin{equation}\label{case1add}
\begin{aligned}
\displaystyle A&=-384\rho\beta_2k_2^2-36\rho \beta_2k_2^2\sin^2\Big(\frac{\sqrt{7\beta_2}k_2}{k_1}(\xi+A_1)\Big)
+252\rho k_2^2\beta_2\cosh^2(\sqrt{\beta_2}(\eta+A_2))\\
&\pm 168\rho k_2^2\beta_2\sin\Big(\frac{\sqrt{7\beta_2}k_2}{k_1}(\xi+A_1)\Big)\cosh(\sqrt{\beta_2}(\eta+A_2)),\\
B&=\zeta\Big[\cos\Big(\frac{\sqrt{7\beta_2}k_2}{k_1}(\xi+A_1)\Big)\pm \sqrt{7}\sinh(\sqrt{\beta_2}(\eta+A_2))\Big]^2,
\end{aligned}
\end{equation}
where $A_1$, $A_2$ are arbitrary constants. This is a novel complexiton solution of the KdV equation. To see the solution's behavior we will give its graphs at some fixed times.

For some specific values of the parameters and particular choice of signs;
$$\displaystyle \zeta=-6,\, \rho=1,\, \alpha_2=-4,\, \beta_2=\frac{1}{4},\, \mu_1=1,\, \mu_2=-2,\, k_1=2,\, k_2=1,\, a_0=1/6,\, A_1=A_2=0,$$
we get the solution
\begin{equation}
\displaystyle u(x,t)=\frac{1}{6}+\frac{A}{B},
\end{equation}
where
\begin{equation}
\begin{aligned}
A&=32+3\sin^2\Big(\frac{\sqrt{7}}{2}(x-7t)\Big)+14\sin\Big(\frac{\sqrt{7}}{2}(x-7t)\Big)\cosh\Big(\frac{1}{2}(x-3t)\Big)
\\&-21\cosh^2\Big(\frac{1}{2}(x-3t)\Big),\\
B&=2\Big[\cos\Big(\frac{\sqrt{7}}{2}(x-7t)\Big)-\sqrt{7}\sinh\Big(\frac{1}{2}(x-3t)\Big)\Big]^2.
\end{aligned}
\end{equation}
\newpage
\noindent The graphs of the above solution are given as follows:
\begin{center}
\begin{figure}[h]
\centering
\begin{minipage}[t]{0.2\textwidth}
\includegraphics[width=\textwidth]{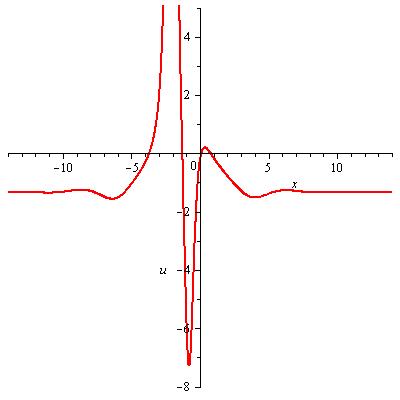}
\caption{t=0}
\end{minipage}%
  \hfill
\begin{minipage}[t]{0.2\textwidth}
\includegraphics[width=\textwidth]{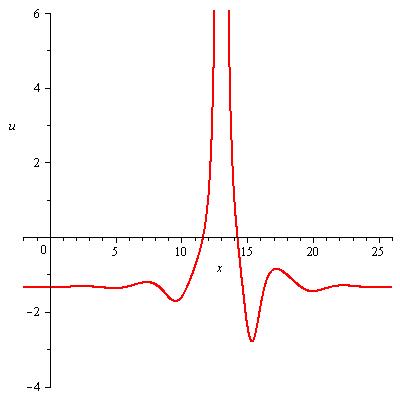}
\caption{t=5}
\end{minipage}
  \hfill
\begin{minipage}[t]{0.2\textwidth}
\includegraphics[width=\textwidth]{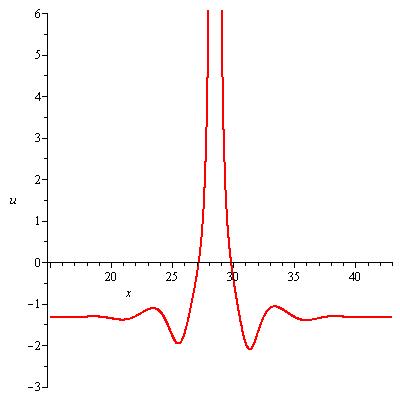}
\caption{t=10}
\end{minipage}%
  \hfill
\begin{minipage}[t]{0.2\textwidth}
\includegraphics[width=\textwidth]{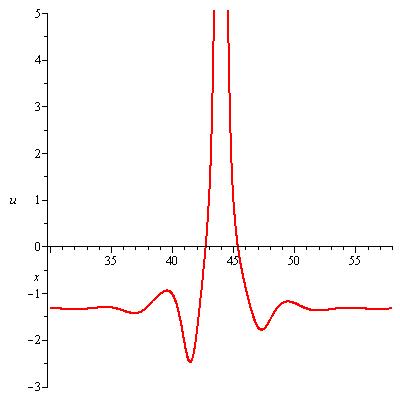}
\caption{t=15}
\end{minipage}
\end{figure}
\end{center}

\noindent \textbf{ii)} \textbf{$\alpha_2>0$, $\beta_2<0$}. The functions $F(\xi)$ and $G(\eta)$ are
\begin{equation*}
F(\xi)= \pm \.{i}\frac{\sqrt{\alpha_2}\mu_2k_1}{7\sqrt{-\beta_2}\mu_1k_2}\sinh\Big(\frac{\sqrt{-7\beta_2}k_2}{k_1}(\xi+A_1) \Big),\quad
G(\eta)=\pm \frac{\sqrt{\alpha_2}}{\sqrt{-\beta_2}}\sin(\sqrt{-\beta_2}(\eta+A_2)),
\end{equation*}
\noindent and the solution is
\begin{equation}
\displaystyle u(x,t)=a_0+\frac{A}{B},
\end{equation}
where
\begin{eqnarray*}
A&=&384\rho\beta_2^2k_2^2-36\rho\beta_2k_2^2\sinh^2\Big(\frac{\sqrt{-7\beta_2}k_2}{k_1}(\xi+A_1) \Big)-252\rho\beta_2k_2^2\sin^2(\sqrt{-\beta_2}(\eta+A_2))\\
&&\pm 168\.{i}\rho\beta_2k_2^2\sinh\Big(\frac{\sqrt{-7\beta_2}k_2}{k_1}(\xi+A_1) \Big)\sin(\sqrt{-\beta_2}(\eta+A_2)),\\
B&=&\zeta\Big[\.{i}\cosh\Big(\frac{\sqrt{-7\beta_2}k_2}{k_1}(\xi+A_1) \Big)\mp \sqrt{7}\cos(\sqrt{-\beta_2}(\eta+A_2))\Big]^2.
\end{eqnarray*}

\noindent If we separate the real and imaginary parts of the above solution we get,
\begin{equation}\label{realimag1}
\displaystyle \mathrm{Re}(u(x,t))=U(x,t)=a_0+\frac{B_1}{C_1},\quad \mathrm{Im}(u(x,t))=V(x,t)=\frac{B_2}{C_1},
\end{equation}
where
\begin{equation}\label{coupledB1B2}
\begin{aligned}
B_1&=12\rho\beta_2k_2^2[-14\cosh^2(\sigma_1)
+3\cosh^4(\sigma_1)+147\cos^4(\sigma_2)-42\cos^2(\sigma_2)\cosh^2(\sigma_1)\\
&-28\sqrt{7}\sin(\sigma_2)\cos(\sigma_2)
\sinh(\sigma_1)\cosh(\sigma_1)+98\cos^2(\sigma_2)],\\
B_2&=\pm 12\rho\beta_2k_2^2[-42\sqrt{7}\cosh(\sigma_1)\cos^3(\sigma_2)+14\sinh(\sigma_1)\cosh^2(\sigma_1)\sin(\sigma_2)\\
&-28\sqrt{7}\cosh(\sigma_1)\cos(\sigma_2)
-98\sinh(\sigma_1)\sin(\sigma_2)\cos^2(\sigma_2)+6\sqrt{7}\cosh^3(\sigma_1)\cos(\sigma_2)],
\end{aligned}
\end{equation}
\noindent and
\begin{equation}\label{coupledC1}
C_1=\zeta[\cosh^2(\sigma_1)+7\cos^2(\sigma_2)]^2,
\end{equation}
with $\displaystyle \sigma_1=\sqrt{-7\beta_2}k_2(\xi+A_1)/k_1$, $\sigma_2=\sqrt{-\beta_2}(\eta+A_2)$. Indeed, the couple $(U(x,t),V(x,t))$
is a novel non-singular solution of the coupled KdV equation (\ref{coupledKdV}).

For the following choice of the parameters; $\displaystyle \zeta=-6, \rho=1, \alpha_2=4, \beta_2=-1, \mu_1=1, \mu_2=1, k_1=2, k_2=1, a_0=1/6, A_1=A_2=0$, we have complexiton solution (\ref{realimag1}) where
\begin{eqnarray*}
\displaystyle
B_1&=& 2[-14\cosh^2(\sqrt{7}(x+33t))
+3\cosh^4(\sqrt{7}(x+33t))+98\cos^2(x+17t)\\
&&-28\sqrt{7}\sinh(\sqrt{7}(x+33t))\cosh(\sqrt{7}(x+33t))\sin(x+17t)\cos(x+17t)
\\&&-42\cos^2(x+17t)\cosh^2(\sqrt{7}(x+33t))+147\cos^4(x+17t)],\\
B_2&=& 4[-14\sqrt{7}\cosh(\sqrt{7}(x+33t))\cos(x+17t)
-21\sqrt{7}\cosh(\sqrt{7}(x+33t))\cos^3(x+17t)\\
&&+7\cosh^2(\sqrt{7}(x+33t))\sinh(\sqrt{7}(x+33t))\sin(x+17t)\\
&&-49\sinh(\sqrt{7}(x+33t))\sin(x+17t)\cos^2(x+17t)\\
&&+3\sqrt{7}\cosh^3(\sqrt{7}(x+33t))\cos(x+17t)],
\end{eqnarray*}
and
\begin{equation*}
C_1=[\cosh^2(\sqrt{7}(x+33t))+7\cos^2(x+17t)]^2.
\end{equation*}
\smallskip
Graphs of the function $U(x,t)$ at some fixed times;
\begin{center}
\begin{figure}[h]
\centering
\begin{minipage}[t]{0.2\textwidth}
\includegraphics[width=\textwidth]{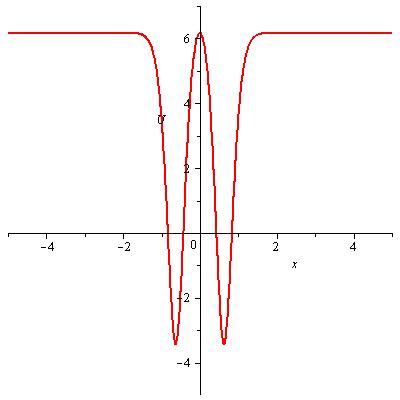}
\caption{t=0}
\end{minipage}%
  \hfill
\begin{minipage}[t]{0.2\textwidth}
\includegraphics[width=\textwidth]{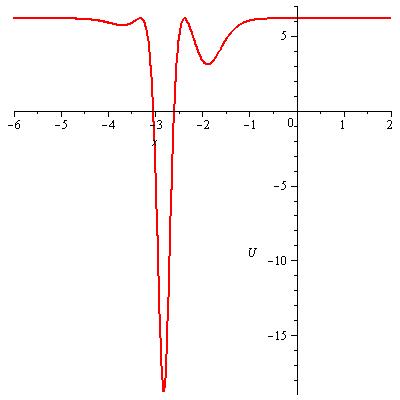}
\caption{t=0.08}
\end{minipage}
  \hfill
\begin{minipage}[t]{0.2\textwidth}
\includegraphics[width=\textwidth]{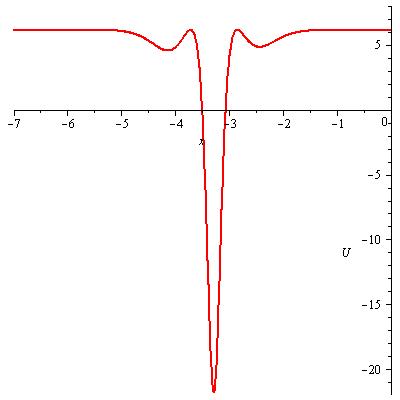}
\caption{t=0.1}
\end{minipage}%
  \hfill
\begin{minipage}[t]{0.2\textwidth}
\includegraphics[width=\textwidth]{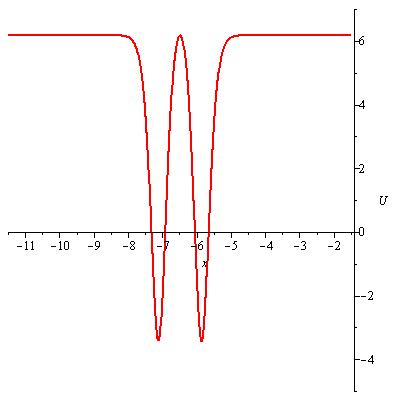}
\caption{t=0.1963}
\end{minipage}
\end{figure}
\end{center}
\smallskip
\noindent And graphs of the function $V(x,t)$ are the followings.
\begin{center}
\begin{figure}[h]
\centering
\begin{minipage}[t]{0.2\textwidth}
\includegraphics[width=\textwidth]{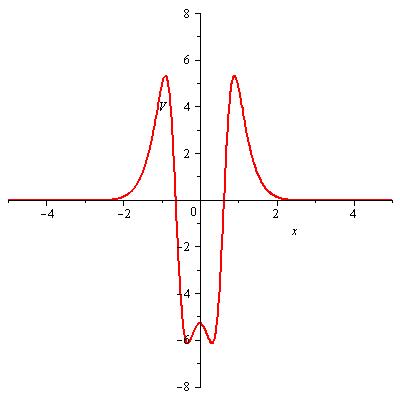}
\caption{t=0}
\end{minipage}%
  \hfill
\begin{minipage}[t]{0.2\textwidth}
\includegraphics[width=\textwidth]{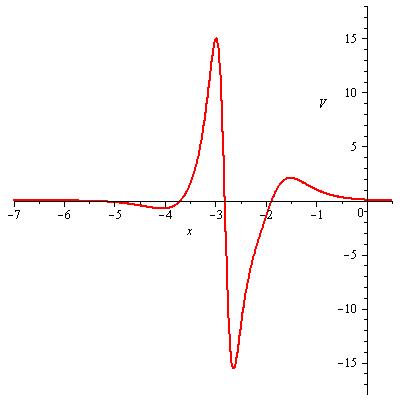}
\caption{t=0.08}
\end{minipage}
  \hfill
\begin{minipage}[t]{0.2\textwidth}
\includegraphics[width=\textwidth]{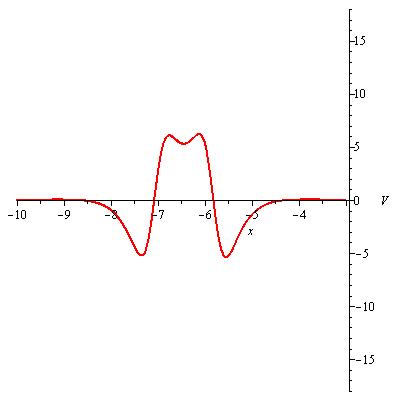}
\caption{t=0.195}
\end{minipage}%
  \hfill
\begin{minipage}[t]{0.2\textwidth}
\includegraphics[width=\textwidth]{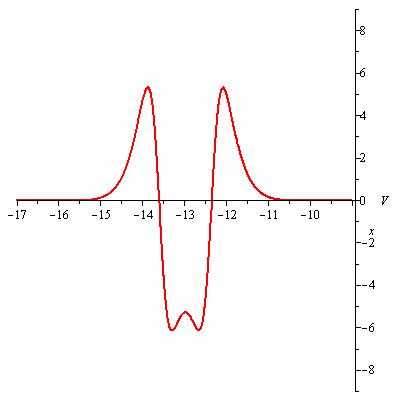}
\caption{t=0.393}
\end{minipage}
\end{figure}
\end{center}

\noindent \textbf{iii)} \textbf{$\alpha_2>0$, $\beta_2>0$}. The functions $F(\xi)$ and $G(\eta)$ are
\begin{equation*}
F(\xi)= \pm \.{i}\frac{\sqrt{\alpha_2}\mu_2k_1}{7\sqrt{\beta_2}\mu_1k_2}\sin\Big(\frac{\sqrt{7\beta_2}k_2}{k_1}(\xi+A_1) \Big),\quad
G(\eta)=\pm \frac{\sqrt{\alpha_2}}{\sqrt{\beta_2}}\sinh(\sqrt{\beta_2}(\eta+A_2)),
\end{equation*}
and the solution is
\begin{equation}
\displaystyle u(x,t)=U(x,t)+\.{i} V(x,t),
\end{equation}
where $\displaystyle U(x,t)=a_0+\frac{B_1}{C_1}$ and $\displaystyle V(x,t)=\frac{B_2}{C_1}$ with
\begin{eqnarray*}
B_1&=&12\rho\beta_2k_2^2[-14\cos^2(\sigma_1)
+3\cos^4(\sigma_1)+147\cosh^4(\sigma_2)-42\cosh^2(\sigma_2)\cos^2(\sigma_1)\\
&&+28\sqrt{7}\sinh(\sigma_2)\cosh(\sigma_2)
\sin(\sigma_1)\cos(\sigma_1)+98\cosh^2(\sigma_2)],\\
B_2&=&\pm 12\rho\beta_2k_2^2[-42\sqrt{7}\cos(\sigma_1)\cosh^3(\sigma_2)-14\sin(\sigma_1)\cos^2(\sigma_1)\sinh(\sigma_2)\\
&&-28\sqrt{7}\cos(\sigma_1)\cosh(\sigma_2)
+98\sin(\sigma_1)\sinh(\sigma_2)\cosh^2(\sigma_2)+6\sqrt{7}\cos^3(\sigma_1)\cosh(\sigma_2)],
\end{eqnarray*}
and
\begin{equation*}
C_1=\zeta[\cos^2(\sigma_1)+ 7\cosh^2(\sigma_2)]^2,
\end{equation*}
with $\displaystyle \sigma_1=\sqrt{7\beta_2}k_2(\xi+A_1)/k_1$, $\sigma_2=\sqrt{\beta_2}(\eta+A_2)$. Hence, we obtain another
new non-singular solution $(U(x,t),V(x,t))$ of the coupled KdV equation (\ref{coupledKdV}).

\noindent For a set of specific values and choice of signs; $\displaystyle \zeta=-6, \rho=1, \alpha_2=4, \beta_2=1, \mu_1=1, \mu_2=1, k_1=2, k_2=1, a_0=1/6, A_1=A_2=0$, we have complexiton solutions of the form (\ref{realimag1}) where
\begin{eqnarray*}
\displaystyle
B_1&=& 2[14\cos^2(\sqrt{7}(x-31t))
-3\cos^4(\sqrt{7}(x-31t))-98\cosh^2(x-15t)\\
&&-28\sqrt{7}\sin(\sqrt{7}(x-31t))\cos(\sqrt{7}(x-31t))\sinh(x-15t)\cosh(x-15t)
\\&&+42\cosh^2(x-15t)\cos^2(\sqrt{7}(x-31t))-147\cosh^4(x-15t)],\\
B_2&=& 4[14\sqrt{7}\cos(\sqrt{7}(x-31t))\cos(x-15t)
+21\sqrt{7}\cos(\sqrt{7}(x-31t))\cosh^3(x-15t)\\
&&+7\cos^2(\sqrt{7}(x-31t))\sin(\sqrt{7}(x-31t))\sinh(x-15t)\\
&&-49\sin(\sqrt{7}(x-31t))\sinh(x-15t)\cosh^2(x-15t)\\
&&-3\sqrt{7}\cos^3(\sqrt{7}(x-31t))\cosh(x-15t)],
\end{eqnarray*}
and
\begin{equation*}
C_1=[\cos^2(\sqrt{7}(x-31t))+7\cosh^2(x-15t)]^2.
\end{equation*}
Let us illustrate the above solutions $U(x,t)$ and $V(x,t)$ at some fixed times. The graphs of $U(x,t)$ are
\begin{center}
\begin{figure}[h]
\centering
\begin{minipage}[t]{0.2\textwidth}
\includegraphics[width=\textwidth]{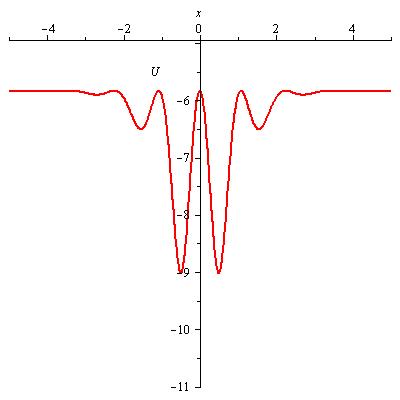}
\caption{t=0}
\end{minipage}%
  \hfill
\begin{minipage}[t]{0.2\textwidth}
\includegraphics[width=\textwidth]{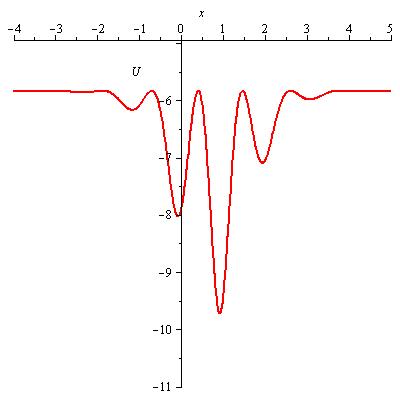}
\caption{t=0.05}
\end{minipage}
  \hfill
\begin{minipage}[t]{0.2\textwidth}
\includegraphics[width=\textwidth]{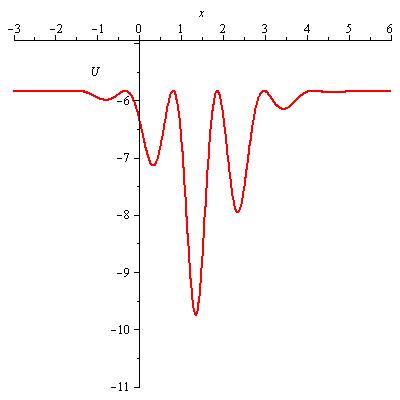}
\caption{t=0.1}
\end{minipage}%
  \hfill
\begin{minipage}[t]{0.2\textwidth}
\includegraphics[width=\textwidth]{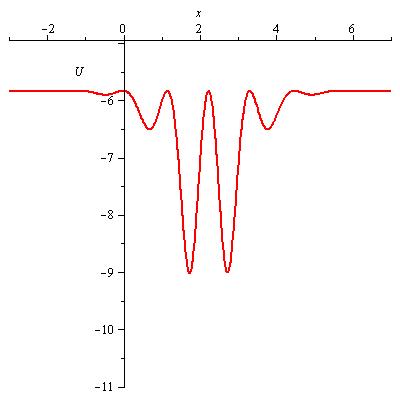}
\caption{t=0.148}
\end{minipage}
\end{figure}
\end{center}
\newpage
\noindent And graphs of the function $V(x,t)$ are the followings.
\begin{center}
\begin{figure}[h]
\centering
\begin{minipage}[t]{0.2\textwidth}
\includegraphics[width=\textwidth]{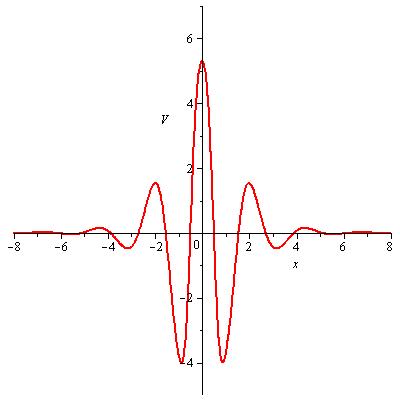}
\caption{t=0}
\end{minipage}%
  \hfill
\begin{minipage}[t]{0.2\textwidth}
\includegraphics[width=\textwidth]{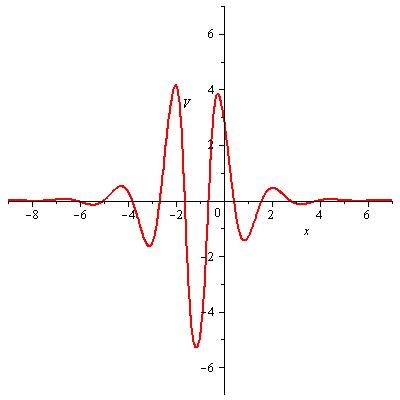}
\caption{t=0.07}
\end{minipage}
  \hfill
\begin{minipage}[t]{0.2\textwidth}
\includegraphics[width=\textwidth]{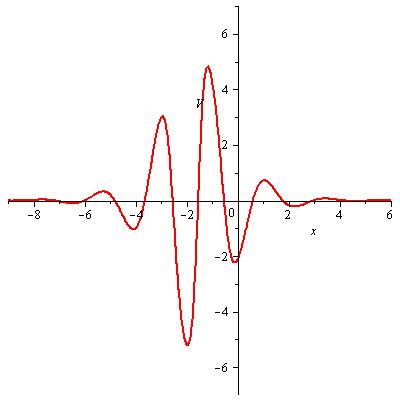}
\caption{t=0.1}
\end{minipage}%
  \hfill
\begin{minipage}[t]{0.2\textwidth}
\includegraphics[width=\textwidth]{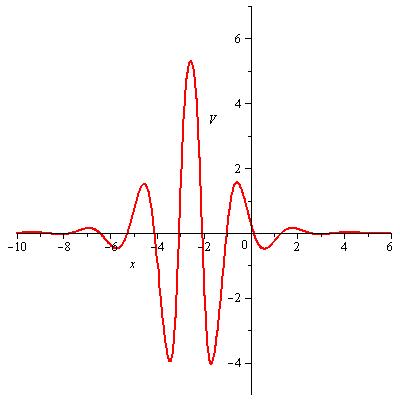}
\caption{t=0.149}
\end{minipage}
\end{figure}
\end{center}

\noindent \textbf{iv)} \textbf{$\alpha_2<0$, $\beta_2<0$}. Here the functions $F(\xi)$ and $G(\eta)$ are
\begin{equation*}
F(\xi)= \pm \frac{\sqrt{-\alpha_2}\mu_2k_1}{7\sqrt{-\beta_2}\mu_1k_2}\sinh\Big(\frac{\sqrt{-7\beta_2}k_2}{k_1}(\xi+A_1) \Big),\quad
G(\eta)=\pm \.{i} \frac{\sqrt{-\alpha_2}}{\sqrt{-\beta_2}}\cos(\sqrt{-\beta_2}(\eta+A_2)),
\end{equation*}
and the solution is
\begin{equation}
\displaystyle u(x,t)=U(x,t)+\.{i} V(x,t),
\end{equation}
where $\displaystyle U(x,t)=a_0+\frac{B_1}{C_1}$ and $\displaystyle V(x,t)=\frac{B_2}{C_1}$ with
\begin{eqnarray*}
B_1&=&12\rho\beta_2k_2^2[-14\cosh^2(\sigma_1)
+3\cosh^4(\sigma_1)+147\sin^4(\sigma_2)-42\sin^2(\sigma_2)\cosh^2(\sigma_1)\\
&&+28\sqrt{7}\sin(\sigma_2)\cos(\sigma_2)
\sinh(\sigma_1)\cos(\sigma_2)+98\sin^2(\sigma_2)],\\
B_2&=&\pm 12\rho\beta_2k_2^2[-42\sqrt{7}\cosh(\sigma_1)\sin^3(\sigma_2)-14\sinh(\sigma_1)\cosh^2(\sigma_1)\cos(\sigma_2)\\
&&-28\sqrt{7}\cosh(\sigma_1)\sin(\sigma_2)
+98\sinh(\sigma_1)\cos(\sigma_2)\sin^2(\sigma_2)+6\sqrt{7}\cosh^3(\sigma_1)\sin(\sigma_2)],
\end{eqnarray*}
and
\begin{equation*}
C_1=\zeta[\cosh^2(\sigma_1)+ 7\sin^2(\sigma_2)]^2,
\end{equation*}
where $\displaystyle \sigma_1=\sqrt{-7\beta_2}k_2(\xi+A_1)/k_1$, $\sigma_2=\sqrt{-\beta_2}(\eta+A_2)$. The above solution
is also a new non-singular solution $(U(x,t),V(x,t))$ of the coupled KdV equation (\ref{coupledKdV}).\\

\noindent Similar solutions can also be obtained with the following set of conditions:

\noindent \textbf{Set 1:}\, $\gamma_1=\gamma_2=a_1=a_2=\kappa_1=\mu_0=0$, and
\begin{eqnarray*}
&&\displaystyle a_3=\frac{4\rho \mu_1^2\beta_2^2k_2^4}{3k_1^2\zeta},\quad a_4=\frac{8\rho\mu_1\mu_2k_2^3\beta_2^2}{k_1\zeta}, \quad a_5=-\frac{4\rho\mu_2^2\beta_2^2k_2^2}{\zeta},\quad \kappa_2=\frac{32\rho k_2^2\mu_2^2\beta_2\alpha_2}{\zeta},\\
&& \alpha_1=-\frac{3\alpha_2\mu_2^2}{\mu_1^2},\quad \beta_1=-\frac{\beta_2 k_2^2}{3k_1^2},\quad c_1=-\frac{4}{3}\rho\beta_2k_2^2+a_0\zeta,\quad c_2=-4\rho\beta_2k_2^2+a_0\zeta.
\end{eqnarray*}
\noindent \textbf{Set 2:}\, We have one more set of conditions; $\gamma_1=\gamma_2=a_1=a_2=\kappa_1=\mu_0=0$, and
\begin{eqnarray*}
&&\displaystyle a_3=-\frac{36\rho k_2^4\beta_2^2\mu_1^2}{k_1^2\zeta} ,\quad a_4=\frac{72\rho\mu_1\mu_2\beta_2^2k_2^3}{k_1\zeta}, \quad a_5=\frac{12\rho\mu_2^2\beta_2^2k_2^2}{\zeta},\quad \kappa_2=\frac{32\rho k_2^2\mu_2^2\beta_2\alpha_2}{\zeta},\\
&& \alpha_1=-\frac{\alpha_2\mu_2^2}{3\mu_1^2},\quad \beta_1=-\frac{3\beta_2 k_2^2}{k_1^2},\quad c_1=12\rho\beta_2k_2^2+a_0\zeta,\quad c_2=4\rho\beta_2k_2^2+a_0\zeta.
\end{eqnarray*}

\bigskip
\noindent \textbf{Case 2.} Let $\gamma_1=\gamma_2=\mu_0=a_1=a_2=a_3=a_5=\kappa_1=0$,
\begin{eqnarray*}
&& a_4=\frac{24\rho\mu_1\mu_2\beta_2^2k_2^3}{k_1\zeta}, \quad \kappa_2=\frac{24\rho\mu_2^2\beta_2 k_2^2\alpha_2}{\zeta}, \quad \alpha_1=-\frac{\alpha_2\mu_2^2}{\mu_1^2}\\
&& \beta_1=-\frac{\beta_2k_2^2}{k_1^2}, \quad c_1=2\rho\beta_2k_2^2+a_0\zeta, \quad c_2=-2\rho\beta_2k_2^2+a_0\zeta.
\end{eqnarray*}

\noindent \textbf{i)} \textbf{$\alpha_2<0$, $\beta_2>0$}. Similar to Case 1., the only case that we have real-valued solutions is when $\alpha_2<0$ and $\beta_2>0$. In this case, we obtain the functions $F(\xi)$ and $G(\eta)$ from (\ref{ODEs}) as
\begin{equation*}
\displaystyle F(\xi)=\pm \frac{\sqrt{-\alpha_2}k_1\mu_2}{\sqrt{\beta_2}\mu_1k_2}\sin\Big( \frac{\sqrt{\beta_2}k_2}{k_1}(\xi+A_1)\Big),\quad
G(\eta)=\pm \frac{\sqrt{-\alpha_2}}{\sqrt{\beta_2}}\cosh(\sqrt{\beta_2}(\eta+A_2)),
\end{equation*}
so the solution becomes
\begin{equation}\label{case2soln}
\displaystyle u(x,t)=a_0+\frac{-24\rho\beta_2k_2^2 \pm 24\rho\beta_2k_2^2 \sin\Big( \frac{\sqrt{\beta_2}k_2}{k_1}(\xi+A_1)\Big)\cosh(\sqrt{\beta_2}(\eta+A_2))}
{\zeta\Big[\cos\Big( \frac{\sqrt{\beta_2}k_2}{k_1}(\xi+A_1)\Big)\pm\sinh(\sqrt{\beta_2}(\eta+A_2))\Big]^2},
\end{equation}
where $\displaystyle \delta_1=24\rho \mu_2^2\beta_2k_2^2/\zeta$ and $A_1$, $A_2$ are arbitrary constants.
Indeed, by some change of variables this solution can be reduced to the complexiton solution of the KdV equation in  \cite{MaComplexiton1}-\cite{MaComplexiton3}.

\noindent For the following choice of the parameters and signs;
$$\displaystyle \zeta=-6, \rho=1, \alpha_2=-4, \beta_2=\frac{1}{4}, \mu_1=1, \mu_2=-1, k_1=-1, k_2=2, a_0=\frac{1}{6}, A_1=A_2=0,$$
we get the solution
\begin{equation}\label{case2ex}
\displaystyle u(x,t)=\frac{1}{6}+\frac{4-4\sin(-x+t)\cosh(x+3t)}{[\cos(-x+t)-\sinh(x+3t)]^2}.
\end{equation}
\noindent The graphs of this solution are given as follows:
\begin{center}
\begin{figure}[h]
\centering
\begin{minipage}[t]{0.2\textwidth}
\includegraphics[width=\textwidth]{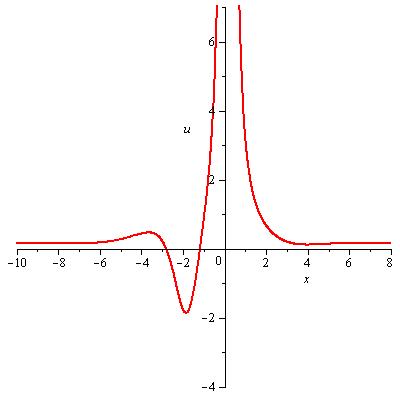}
\caption{t=0}
\end{minipage}%
  \hfill
\begin{minipage}[t]{0.2\textwidth}
\includegraphics[width=\textwidth]{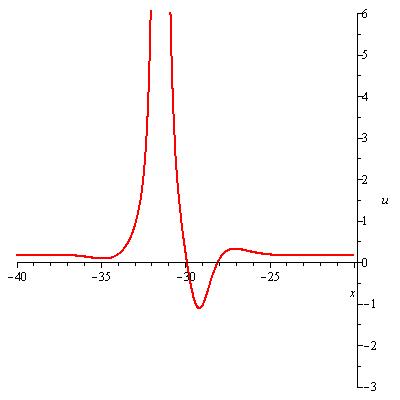}
\caption{t=10}
\end{minipage}
  \hfill
\begin{minipage}[t]{0.2\textwidth}
\includegraphics[width=\textwidth]{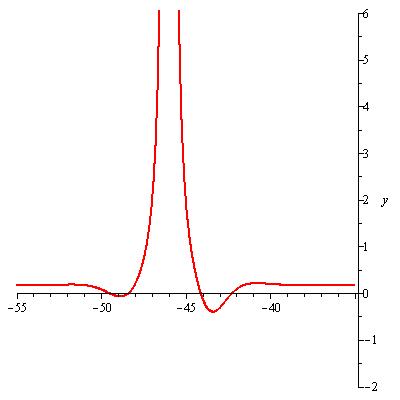}
\caption{t=15}
\end{minipage}%
  \hfill
\begin{minipage}[t]{0.2\textwidth}
\includegraphics[width=\textwidth]{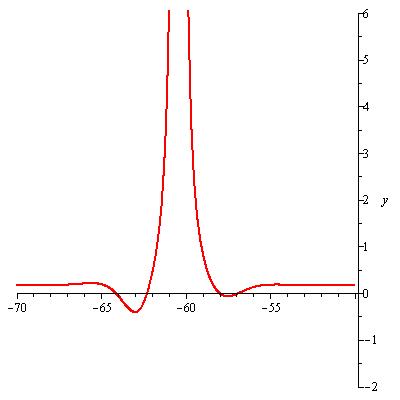}
\caption{t=20}
\end{minipage}
\end{figure}
\end{center}

\noindent \textbf{ii)} \textbf{$\alpha_2>0$, $\beta_2<0$}. In this case, the functions $F(\xi)$ and $G(\eta)$ become
\begin{equation*}
\displaystyle F(\xi)= \pm \.{i} \frac{\sqrt{\alpha_2}\mu_2k_1}{\sqrt{-\beta_2}\mu_1k_2}\sinh\Big(\frac{\sqrt{-\beta_2}k_2}{k_1}(\xi+A_1) \Big),\quad
G(\eta)=\pm  \frac{\sqrt{\alpha_2}}{\sqrt{-\beta_2}}\sin(\sqrt{-\beta_2}(\eta+A_2)),
\end{equation*}
and the solution is
\begin{equation}
\displaystyle u(x,t)=U(x,t)+\.{i} V(x,t),
\end{equation}
where
\begin{equation}\label{case2coupled}
\displaystyle U(x,t)=a_0+\frac{A}{C},\quad V(x,t)=\frac{B}{C},
\end{equation}
where
\begin{equation}\label{case2coupledterms}
\begin{aligned}
A&=a_0+24\rho\beta_2k_2^2[\cos^2(\sigma_1)-\cosh^2(\sigma_2)-2\sin(\sigma_1)\cos(\sigma_1)\sinh(\sigma_2)\cosh(\sigma_2)],\\
B&=24\rho\beta_2k_2^2[\sin(\sigma_1)\sinh(\sigma_2)\cosh^2(\sigma_2)-2\cos(\sigma_1)\cosh(\sigma_2)
-\sinh(\sigma_2)\sin(\sigma_1)\cos^2(\sigma_1)],\\
C&=\zeta[\cosh^2(\sigma_2)+\cos^2(\sigma_1)]^2,
\end{aligned}
\end{equation}
with $\sigma_1=\sqrt{-\beta_2}(\eta+A_2)$ and $\displaystyle \sigma_2=\sqrt{-\beta_2}k_2(\xi+A_1)/k_1$, and $A_1$, $A_2$ are arbitrary constants.

\noindent For a set of specific values and choice of signs; $\displaystyle \zeta=-6, \rho=1, \alpha_2=4, \beta_2=-\frac{1}{4}, \mu_1=1, \mu_2=-1, k_1=-1, k_2=2, a_0=1/6, A_1=A_2=0$, we have non-singular complexiton solutions of the form (\ref{case2coupled}) where the terms $A$, $B$, and $C$ in (\ref{case2coupledterms}) becomes
\begin{eqnarray*}
\displaystyle
A&=& 4[\cos^2(-x+t)-\cosh^2(x+3t)+2\sinh(x+3t)\cosh(x+3t)\sin(-x+t)\cos(-x+t)],\\
B&=& 4[\sinh(x+3t)\sin(-x+t)\cos^2(-x+t)-\sin(-x+t)\sinh(x+3t)\cosh^2(x+3t)\\
&&-2\cos(-x+t)\cosh(x+3t)],
\end{eqnarray*}
and
\begin{equation*}
C=[\cosh^2(x+3t)+\cos^2(-x+t)]^2.
\end{equation*}
Firstly, let us present the graphs of the solution $U(x,t)$,
\begin{center}
\begin{figure}[h]
\centering
\begin{minipage}[t]{0.2\textwidth}
\includegraphics[width=\textwidth]{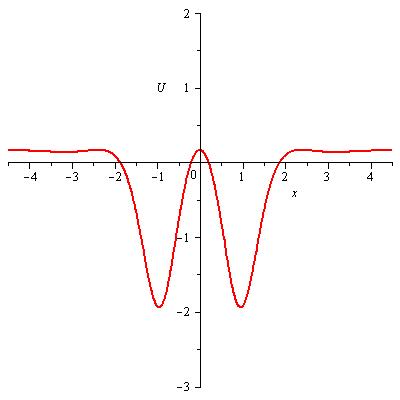}
\caption{t=0}
\end{minipage}%
  \hfill
\begin{minipage}[t]{0.2\textwidth}
\includegraphics[width=\textwidth]{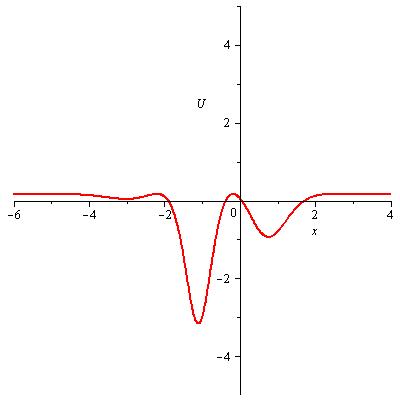}
\caption{t=0.2}
\end{minipage}
  \hfill
\begin{minipage}[t]{0.2\textwidth}
\includegraphics[width=\textwidth]{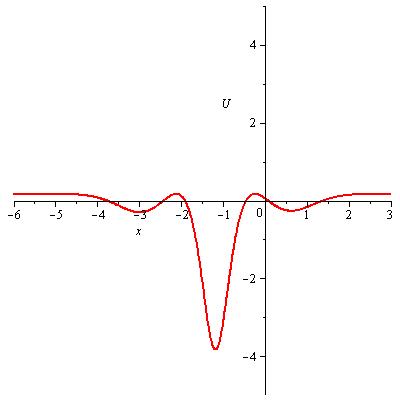}
\caption{t=0.4}
\end{minipage}%
  \hfill
\begin{minipage}[t]{0.2\textwidth}
\includegraphics[width=\textwidth]{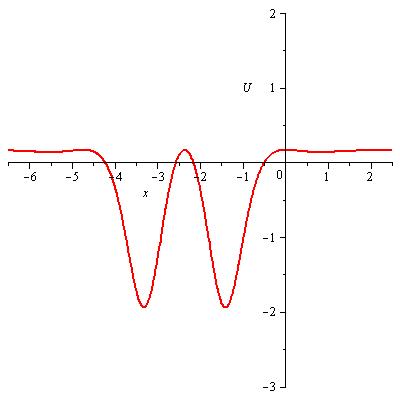}
\caption{t=0.785}
\end{minipage}
\end{figure}
\end{center}
\newpage
\noindent The graphs of the solution $V(x,t)$ are
\begin{center}
\begin{figure}[h]
\centering
\begin{minipage}[t]{0.2\textwidth}
\includegraphics[width=\textwidth]{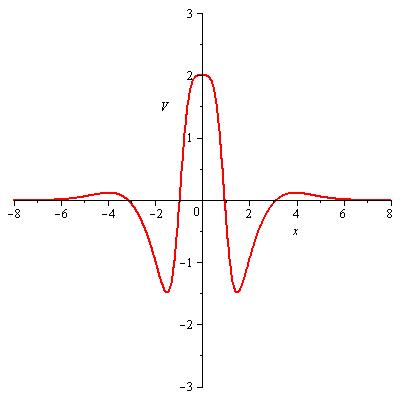}
\caption{t=0}
\end{minipage}%
  \hfill
\begin{minipage}[t]{0.2\textwidth}
\includegraphics[width=\textwidth]{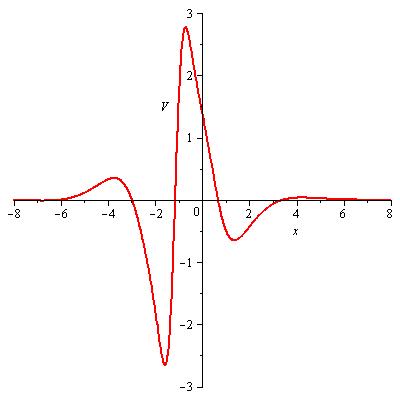}
\caption{t=0.3}
\end{minipage}
  \hfill
\begin{minipage}[t]{0.2\textwidth}
\includegraphics[width=\textwidth]{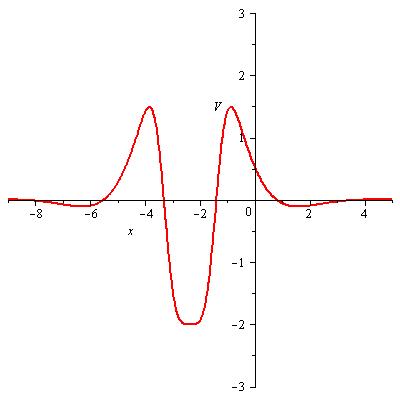}
\caption{t=0.785}
\end{minipage}%
  \hfill
\begin{minipage}[t]{0.2\textwidth}
\includegraphics[width=\textwidth]{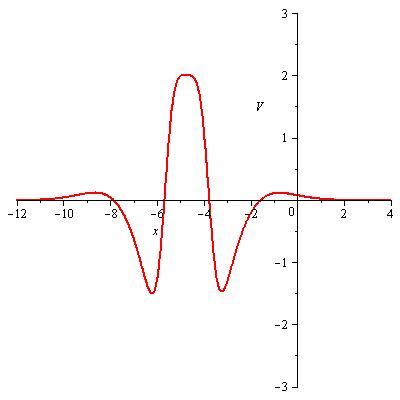}
\caption{t=1.575}
\end{minipage}
\end{figure}
\end{center}
\noindent \textbf{iii)} \textbf{$\alpha_2>0$, $\beta_2>0$}. The functions $F(\xi)$ and $G(\eta)$ are
\begin{equation*}
\displaystyle
F(\xi)= \pm \.{i} \frac{\sqrt{\alpha_2}\mu_2k_1}{\sqrt{\beta_2}\mu_1k_2}\sin\Big(\frac{\sqrt{\beta_2}k_2}{k_1}(\xi+A_1) \Big),\quad
G(\eta)=\pm  \frac{\sqrt{\alpha_2}}{\sqrt{\beta_2}}\sinh(\sqrt{\beta_2}(\eta+A_2)),
\end{equation*}
and the solution is
\begin{equation}
\displaystyle u(x,t)=U(x,t)+\.{i} V(x,t).
\end{equation}
Here we have
\begin{equation}
\displaystyle U(x,t)=a_0+\frac{A}{C},\quad V(x,t)=\frac{B}{C},
\end{equation}
where
\begin{equation}
\begin{aligned}
A&=a_0+24\rho\beta_2k_2^2[\cosh^2(\sigma_1)-\cos^2(\sigma_2)+2\sinh(\sigma_1)\cosh(\sigma_1)\sin(\sigma_2)\cos(\sigma_2)],\\
B&=\pm 24\rho\beta_2k_2^2[\sin(\sigma_2)\sinh(\sigma_1)\cosh^2(\sigma_1)-\sinh(\sigma_1)\sin(\sigma_2)\cos^2(\sigma_2)-2\cosh(\sigma_1)\cos(\sigma_2)], \\
C&=\zeta[\cosh^2(\sigma_1)+\cos^2(\sigma_2)]^2,
\end{aligned}
\end{equation}
with $\sigma_1=\sqrt{\beta_2}(\eta+A_2)$ and $\displaystyle \sigma_2=\sqrt{\beta_2}k_2(\xi+A_1)/k_1$, and $A_1$, $A_2$ are arbitrary constants. The couple $(U(x,t),V(x,t))$ is another
solution of (\ref{coupledKdV}).

\noindent \textbf{iv)} \textbf{$\alpha_2<0$, $\beta_2<0$}. The functions $F(\xi)$ and $G(\eta)$ are
\begin{equation*}
\displaystyle
F(\xi)= \pm  \frac{\sqrt{-\alpha_2}\mu_2k_1}{\sqrt{-\beta_2}\mu_1k_2}\sinh\Big(\frac{\sqrt{-\beta_2}k_2}{k_1}(\xi+A_1) \Big),\quad
G(\eta)=\pm  \.{i}\frac{\sqrt{-\alpha_2}}{\sqrt{-\beta_2}}\cos(\sqrt{-\beta_2}(\eta+A_2)),
\end{equation*}
and the solution is
\begin{equation}
\displaystyle u(x,t)=U(x,t)+\.{i} V(x,t),
\end{equation}
where
\begin{equation}
\displaystyle U(x,t)=a_0+\frac{A}{C},\quad V(x,t)=\frac{B}{C},
\end{equation}
with
\begin{equation}
\begin{aligned}
A&=a_0+24\rho\beta_2k_2^2[\sin^2(\sigma_1)-\cosh^2(\sigma_2)+2\sin(\sigma_1)\cos(\sigma_1)\sinh(\sigma_2)\cosh(\sigma_2)],\\
B&=\pm 24\rho\beta_2k_2^2[\sinh(\sigma_2)\cos(\sigma_1)\sin^2(\sigma_1)-\cos(\sigma_1)\sinh(\sigma_2)\cosh^2(\sigma_2)-2\sin(\sigma_1)\cosh(\sigma_2)], \\
C&=\zeta[\cosh^2(\sigma_2)+\sin^2(\sigma_1)]^2,
\end{aligned}
\end{equation}
where $\sigma_1=\sqrt{-\beta_2}(\eta+A_2)$ and $\displaystyle \sigma_2=\sqrt{-\beta_2}k_2(\xi+A_1)/k_1$, and $A_1$, $A_2$ are arbitrary constants. The couple $(U(x,t),V(x,t))$ is a solution of (\ref{coupledKdV}).

\section{Asymptotic Behaviors of the Solutions}

\noindent Both of the complexiton solutions obtained in Case 1. that is the solution (\ref{case1soln}) with (\ref{case1add}) and the solution (\ref{case2soln}) found in Case 2. are singular solutions. For any parameters in these solutions, obviously the denominators of them vanish at some points at a fixed time. The solution (\ref{case1soln}) with (\ref{case1add}) approaches to a constant $\displaystyle a_0+252\rho k_2^2\beta_2/7\zeta$
as $x\rightarrow \pm \infty$ and similarly as $t \rightarrow \pm \infty$. Other solution (\ref{case2soln}) obtained in Case 2. approaches to $a_0$ as $x\rightarrow \pm \infty$ and as $t \rightarrow \pm \infty$. Even approaching to a constant is one of the main properties of solitary wave solutions, it is clear that these solutions are not solitary wave solutions since they have blowing-ups which can also be realized through the graphs given. The graphs also show that because of the trigonometric functions both of the solutions tend to be periodic but by hyperbolic functions their periodicity fades away.

Other solutions that are complex-valued solutions of the KdV equation give real-valued non-singular solutions of the coupled KdV equation (\ref{coupledKdV}). For instance, in Case 1. we get the solution (\ref{realimag1}) as a couple $(U(x,t),V(x,t))$ with (\ref{coupledB1B2}) and (\ref{coupledC1}). Notice that while the solution $U(x,t)$ approaches to $a_0+36\rho \beta_2 k_2^2/\zeta$ as $x\rightarrow \pm \infty$, and $t\rightarrow \pm \infty$, the solution $V(x,t)$ approaches to zero. If the graphs of the solutions $U(x,t)$ and $V(x,t)$ are analyzed, one can realize that the waves defined by the solutions change their forms as time changes but at some point they return to their original shapes.

\section{Conclusion}

We presented a new approach to double-sub equation method which is a practical and an appropriate method for symbolic computation in MAPLE or Mathematica to explore novel solutions of nonlinear partial differential equations. Specifically, we applied this method on the $(1+1)$-dimensional KdV equation and obtained periodic and solitary wave solutions, and notably complexiton solutions depending on two independent variables which are novel solutions to the best of our knowledge. In addition to that application of this approach produced new complexiton solutions of the coupled KdV equation.

This method can be applied to many other nonlinear integrable and non-integrable equations. In this study, we used a $(1+1)$-dimensional equation but by introducing an additional function and a first order ODE satisfied by this new function we can also apply this method to $(2+1)$-dimensional or even higher order partial differential equations. Indeed, it is worth to generalize this approach and study on a unification of many of the solution methods in the literature which may also produce interesting solutions of well-known partial differential equations.

\section{Acknowledgment}
  This work is partially supported by the Scientific
and Technological Research Council of Turkey (T\"{U}B\.{I}TAK).


\begin{thebibliography}{}

\bibitem{Bo} Boussinesq J.V., Essai sur la th\'{e}orie des eaux courantes, M\'{e}moires pr\'{e}sent\'{e}s par divers savants \`{a} l'Acad. des
Sci. Inst. Nat. France, XXIII(1): 1--680 (1877).

\bibitem{KdV} Korteweg D.J., de Vries G., {\it On the change of form of long waves advancing in a rectangular canal and,
on a new type of long stationary waves}, Philosophical Magazine {\bf 39}, 240, 442--443 (1895).

\bibitem{Lax} Lax P.D., {\it Integrals of non-linear equations of evolution and solitary waves}, Commun. Pure Appl. Math. {\bf 21}, 467--490
(1968).

\bibitem{Olver} Olver P.J., {\it Evolution equations possessing infinitely many symmetries}, J. Math. Phys. {\bf 18}, 1212 (1977).

\bibitem{WahEst} Wahlquist H.D., Estabrook F.B., {\it B\"{a}cklund transformation for solutions of the Korteweg-de Vries equation},
Phys. Rev. Lett. {\bf 31}, 1386 (1973).



\bibitem{GGKM} Gardner C.S., Greene J.M., Kruskal M.D., and Miura R.M., {\it Method
for solving the Korteweg-de Vries equation}, Phys. Rev. Lett. {\bf 19}, 1095--1097 (1967).

\bibitem{AbSe} Ablowitz M., Segur H., Solitons and the Inverse Scattering Transform, SIAM, Philadelphia, (1981).

\bibitem{AbCl} Ablowitz M., Clarkson P., Solitons, Nonlinear Evolution Equations and Inverse Scattering, Cambridge University Press, Cambridge, (1991).

\bibitem{HirotaKdV} Hirota R., {\it Exact solution of the Korteweg-de Vries equation for multiple collisions of solitons}, Phys. Rev. Lett. {\bf 27}, 1192 (1971).

\bibitem{Hirota} Hirota R., The Direct Method in Soliton Theory, Cambridge University Press, Cambridge (2004).

\bibitem{pekcan} Pekcan A., {\it Solutions of the Extended
Kadomtsev-Petviashvili-Boussinesq Equation by the Hirota Direct
Method}, Journal of Nonlinear Math. Phys. {\bf 16}, Issue:2, 127--139 (2009).

\bibitem{hietarinta} Hietarinta J., {\it Searching for integrable PDE's by testing Hirota's three-soliton condition}, Proceedings
of the 1991 International Symposium on Symbolic and Algebraic Computation, ISSAC'91, Stephen M.
Watt (Association for Computing Machinery, 1991) 295–-300.


\bibitem{WeissTaborGarnevale} Weiss J., Tabor M., and Garnevale G., {\it The Painlev\'{e} property for partial differential equations}, J. Math. Phys. {\bf
    24}, 522--526 (1982).
\bibitem{NewTabZeng} Newell A.C, Tabor M., and Zeng Y.B., {\it A unified approach to Painlev\'{e} expansions}, Physica D {\bf 29}, (1987).

\bibitem{Sah} Sahadevan R., {\it Painlev\'{e} expansion and exact solution for nonlinear evolution equations}, Theor. and Math. Phys. {\bf 99}, Issue: 3, 776--782 (1994).

\bibitem{RSK} Rasinariu C., Sukhatme U., and Khare A., {\it Negaton and positon solutions of the KdV and mKdV hierarchy}, J. Phys. A: Math. and Gen. {\bf 29}, Number 8, 1803--1823 (1996).

\bibitem{MaComplexiton1} Ma W.X., {\it Complexiton solutions to the Korteweg-de Vries equation}, Phys. Lett. A {\bf 301}, 35--44 (2002).
\bibitem{MaComplexiton2} Ma W.X., Maruno K., {\it Complexiton solutions to the Toda lattice equation}, Physica A {\bf 343}, 219--237 (2004).
\bibitem{MaComplexiton3} Ma W.X., {\it Complexiton solutions to integrable equations}, Nonlinear Analysis {\bf 63}, e2461--e2471 (2005).



\bibitem{subODE1} Li X.Z., Wang M.L., {\it A sub-ODE method for finding exact solutions of a generalized
KdV–mKdV equation with high-order nonlinear terms}, Phys. Lett. A {\bf 361}, 115-–118 (2007).

\bibitem{subODE2} Wang M.L., Li X.Z., and Zhang J.L., {\it Sub-ODE method and solitary wave solutions for
higher order nonlinear Schr\"{o}dinger equation}, Phys. Lett. A {\bf 363}, 96-–101 (2007).

\bibitem{subODE3} Chen H.T, Zhang H.Q., {\it New double periodic and multiple soliton solutions of the generalized (2+1)-dimensional Boussinesq equation}, Chaos Solitons Fract. {\bf 20}, 765-–769 (2004).

\bibitem{chen-yang-ma} Chen H.T., Yang S.H., and Ma W.X., {\it Double sub-equation method for complexiton solutions of nonlinear partial differential equations}, App. Math. and Comp. {\bf 219}, 4775--4781 (2013).

\bibitem{liu-lin-sun} Liu H.Z., Lin S., and Sun X.Q., {\it Comment on: "Double sub-equation method for complexiton solutions of nonlinear partial differential equations"}, App. Math. and Comp. {\bf 246}, 597--598 (2014).

\bibitem{HirotaSat} Hirota R., Satsuma J., {\it Soliton solutions of a coupled Korteweg-de Vries equation}, Phys. Lett. A {\bf 85}, 407--408 (1981).

\bibitem{DrinfeldSokolov} Drinfeld V.G., Sokolov V.V., {\it Lie algebras and equations of Korteweg-de Vries type}, Sov. J. Math. {\bf 30}, 1975--2036 (1985).

\bibitem{Fuch} Fuchssteiner B., {\it The Lie algebra structure of degenerate Hamiltonian and bi-Hamiltonian systems}, Prog. Theor. Phys. {\bf 68}, 1082--1104 (1982).

\bibitem{NO} Nutku Y., O\~{g}uz O., {\it Bi-Hamiltonian structure of a pair of coupled KdV equations}, Nuovo Cimento Soc. Ital. Fis. B {\bf 105}, 1381--1383 (1990).

\bibitem{MGAP1} G\"{u}rses M., Pekcan A., {\it Traveling wave equations of degenerate coupled KdV equation}, J. of Math. Phys.  {\bf 55}, 091501 (2014).

\bibitem{MGAP2} G\"{u}rses M., Pekcan A., {\it Traveling wave solutions of degenerate coupled multi-KdV equations},  arXiv:1506.02362, (2015).


\bibitem{Hu1} Lou S.Y., Tong B., Hu H.C., and Tang X.Y., {\it Coupled KdV equations derived from two-layer fluids}, J. Phys. A: Math. Gen. {\bf 39}, 513--527 (2006).

\bibitem{Bose} Brazhnyi V.A., Konotop V.V., {\it Stable and unstable vector dark solitons of coupled nonlinear Schr\"{o}dinger equations. Application to two-component
Bose-Einstein condensates}, Phys. Rev. E {\bf 72}, 026616 (2005).

\bibitem{Hu2} Hu H.C., Tang B., and Lou S.Y., {\it Nonsingular positon and complexiton solutions for the coupled KdV system},
Phys. Lett. A {\bf 351}, 403--412 (2006).

\bibitem{YangMao1} Yang J.R., Mao J.J., {\it Painlev\'{e} property and complexiton solutions of a special coupled KdV equation}, Commun. Theor. Phys. {\bf 50},
809--813 (2008).

\end{thebibliography}
\end{document}